\documentclass{article}
\usepackage{spconf,amsmath,graphicx}

\usepackage{color}
\usepackage{placeins}
\usepackage{float}
\usepackage{bm}
\usepackage{tabularx,colortbl}
\usepackage{url}

\usepackage{algorithm}
\usepackage[noend]{algpseudocode}

\usepackage{pgfplots}
\usepackage{amssymb}
\pgfplotsset{compat=1.9}
\usepackage{comment}

\newcommand{\R}{\mathbb{R}}

\definecolor{colmint}{HTML}{8DD3C7}
\definecolor{colyellow}{HTML}{FFFFB3}
\definecolor{colorange}{HTML}{FDB462}
\definecolor{colred}{HTML}{FB8072}
\definecolor{colviolet}{HTML}{BEBADA}
\definecolor{colblue}{HTML}{80B1D3}

\pgfplotsset{
	tick label style = {font = \small},
	label style = {font = \small},
	legend style = {font = \footnotesize}
	}
	
\usetikzlibrary{spy}
\usetikzlibrary{calc}

\newif\ifblackandwhitecycle
\gdef\patternnumber{0}

\pgfkeys{/tikz/.cd,
	zoombox paths/.style={
		draw=orange,
		very thick
	},
	black and white/.is choice,
	black and white/.default=static,
	black and white/static/.style={ 
		draw=white,   
		zoombox paths/.append style={
			draw=white,
			postaction={
				draw=black,
				loosely dashed
			}
		}
	},
	black and white/static/.code={
		\gdef\patternnumber{1}
	},
	black and white/cycle/.code={
		\blackandwhitecycletrue
		\gdef\patternnumber{1}
	},
	black and white pattern/.is choice,
	black and white pattern/0/.style={},
	black and white pattern/1/.style={    
		draw=white,
		postaction={
			draw=black,
			dash pattern=on 2pt off 2pt
		}
	},
	black and white pattern/2/.style={    
		draw=white,
		postaction={
			draw=black,
			dash pattern=on 4pt off 4pt
		}
	},
	black and white pattern/3/.style={    
		draw=white,
		postaction={
			draw=black,
			dash pattern=on 4pt off 4pt on 1pt off 4pt
		}
	},
	black and white pattern/4/.style={    
		draw=white,
		postaction={
			draw=black,
			dash pattern=on 4pt off 2pt on 2 pt off 2pt on 2 pt off 2pt
		}
	},
	zoomboxarray inner gap/.initial=2pt,
	zoomboxarray columns/.initial=2,
	zoomboxarray rows/.initial=2,
	zoomboxarray hscale/.initial=1,
	subfigurename/.initial={},
	figurename/.initial={zoombox},
	zoomboxarray/.style={
		execute at begin picture={
			\begin{scope}[
				spy using outlines={%
					zoombox paths,
					width=\imagewidth  / \pgfkeysvalueof{/tikz/zoomboxarray 
					columns} - (\pgfkeysvalueof{/tikz/zoomboxarray columns} - 
					1) / \pgfkeysvalueof{/tikz/zoomboxarray columns} * 
					\pgfkeysvalueof{/tikz/zoomboxarray inner gap} 
					-\pgflinewidth,
					height=\imageheight*(\pgfkeysvalueof{/tikz/zoomboxarray 
					hscale}) / \pgfkeysvalueof{/tikz/zoomboxarray rows} - 
					(\pgfkeysvalueof{/tikz/zoomboxarray rows} - 1) / 
					\pgfkeysvalueof{/tikz/zoomboxarray rows} * 
					\pgfkeysvalueof{/tikz/zoomboxarray inner gap}-
					\pgflinewidth,
					magnification=3,
					every spy on node/.style={
						zoombox paths
					},
					every spy in node/.style={
						zoombox paths
					}
				}
				]
			},
			execute at end picture={
			\end{scope}

			\gdef\patternnumber{0}
		},
		spymargin/.initial=0.5em,
		zoomboxes xshift/.initial=1,
		zoomboxes right/.code=\pgfkeys{/tikz/zoomboxes xshift=1},
		zoomboxes left/.code=\pgfkeys{/tikz/zoomboxes xshift=-1},
		zoomboxes yshift/.initial=0,
		zoomboxes above/.code={
			\pgfkeys{/tikz/zoomboxes yshift=1},
			\pgfkeys{/tikz/zoomboxes xshift=0}
		},
		zoomboxes below/.code={
			\pgfkeys{/tikz/zoomboxes yshift=-0.8},
			\pgfkeys{/tikz/zoomboxes xshift=0}
		},
		caption margin/.initial=4ex
	},
	adjust caption spacing/.code={},
	image container/.style={
		inner sep=0pt,
		at=(image.north),
		anchor=north,
		adjust caption spacing
	},
	zoomboxes container/.style={
		inner sep=0pt,
		at=(image.north),
		anchor=north,
		name=zoomboxes container,
		xshift=\pgfkeysvalueof{/tikz/zoomboxes 
		xshift}*(\imagewidth+\pgfkeysvalueof{/tikz/spymargin}),
		yshift=\pgfkeysvalueof{/tikz/zoomboxes 
		yshift}*(\imageheight+\pgfkeysvalueof{/tikz/spymargin}+
	\pgfkeysvalueof{/tikz/caption
		 margin}),
		adjust caption spacing
	},
	calculate dimensions/.code={
		\pgfpointdiff{\pgfpointanchor{image}{south west} 
		}{\pgfpointanchor{image}{north east} }
		\pgfgetlastxy{\imagewidth}{\imageheight}
		\global\let\imagewidth=\imagewidth
		\global\let\imageheight=\imageheight
		\gdef\columncount{1}
		\gdef\rowcount{1}
		
	},
	image node/.style={
		inner sep=0pt,
		name=image,
		anchor=south west,
		append after command={
			[calculate dimensions]
			node [image 
			container,subfigurename=\pgfkeysvalueof{/tikz/figurename}-image] 
			{\phantomimage}
			node [zoomboxes 
			container,subfigurename=\pgfkeysvalueof{/tikz/figurename}-zoom] {}
		}
	},
	color code/.style={
		zoombox paths/.append style={draw=#1}
	},
	connect zoomboxes/.style={
		spy connection path={\draw[draw=none,zoombox paths] (tikzspyonnode) -- 
		(tikzspyinnode);}
	},
	help grid code/.code={
		\begin{scope}[
			x={(image.south east)},
			y={(image.north west)},
			font=\footnotesize,
			help lines,
			overlay
			]
			\foreach \x in {0,1,...,9} { 
				\draw(\x/10,0) -- (\x/10,1);
				\node [anchor=north] at (\x/10,0) {0.\x};
			}
			\foreach \y in {0,1,...,9} {
				\draw(0,\y/10) -- (1,\y/10);                        
				\node [anchor=east] at (0,\y/10) {0.\y};
			}
		\end{scope}    
	},
	help grid/.style={
		append after command={
			[help grid code]
		}
	}
}

\newcommand\phantomimage{%
	\phantom{%
		\rule{\imagewidth}{\imageheight}%
	}%
}
\newcommand\zoombox[2][]{
	\begin{scope}[zoombox paths]
		\pgfmathsetmacro\xpos{
			(\columncount-1)*(\imagewidth / \pgfkeysvalueof{/tikz/zoomboxarray 
			columns} + \pgfkeysvalueof{/tikz/zoomboxarray inner gap} / 
			\pgfkeysvalueof{/tikz/zoomboxarray columns} ) + \pgflinewidth
			- 0.5*\imagewidth
		}
		\pgfmathsetmacro\ypos{
			(\rowcount-1)*( \imageheight *(\pgfkeysvalueof{/tikz/zoomboxarray 
			hscale}) / \pgfkeysvalueof{/tikz/zoomboxarray rows} + 
			\pgfkeysvalueof{/tikz/zoomboxarray inner gap} / 
			\pgfkeysvalueof{/tikz/zoomboxarray rows} ) + 0.5*\pgflinewidth
		}
		\edef\dospy{\noexpand\spy [
			#1,
			zoombox paths/.append style={
				black and white pattern=\patternnumber
			},
			every spy on node/.append style={#1},
			x=\imagewidth,
			y=\imageheight 
			] on (#2) in node [anchor=north west] at ($(zoomboxes 
			container.north west)+(\xpos pt,-\ypos pt)$);}
		\dospy
		\pgfmathtruncatemacro\pgfmathresult{ifthenelse(\columncount==
			\pgfkeysvalueof{/tikz/zoomboxarray
		 columns},\rowcount+1,\rowcount)}
		\global\let\rowcount=\pgfmathresult
		\pgfmathtruncatemacro\pgfmathresult{ifthenelse(\columncount==
			\pgfkeysvalueof{/tikz/zoomboxarray
		 columns},1,\columncount+1)}
		\global\let\columncount=\pgfmathresult
		\ifblackandwhitecycle
		\pgfmathtruncatemacro{\newpatternnumber}{\patternnumber+1}
		\global\edef\patternnumber{\newpatternnumber}
		\fi
	\end{scope}
}

\def\x{{\mathbf x}}

\title{Compressing Colour Images with Joint Inpainting and Prediction}

\name{Rahul Mohideen Kaja Mohideen$^{\star}$ \quad Pascal Peter$^{\star}$%
 \quad Tobias Alt$^{\star}$ \quad Joachim Weickert$^{\star}$ \quad Alexander Scheer$^{\star}$
  \thanks{This work has received funding from the European Research Council
          (ERC) under the European Union's Horizon 2020 research and 
          innovation programme (grant agreement no. 741215, ERC Advanced
          Grant INCOVID).}
     }

\address{$^{\star}$ Mathematical Image Analysis Group, Faculty of Mathematics and Computer Science\\
Campus E1.7, Saarland University, 
66041 Saarbr\"ucken, Germany \\
\{rakaja, peter,  alt, weickert, scheer\}@mia.uni-saarland.de\\}

\begin{document}
\ninept
\maketitle

\begin{abstract}
Inpainting-based codecs store sparse, quantised pixel data 
directly and decode by interpolating the discarded image parts. This 
interpolation can be used simultaneously for efficient coding by predicting 
pixel data to be stored. Such joint inpainting and prediction approaches 
yield good results with simple components such as regular grids and Shepard 
interpolation on grey value images, but they lack a dedicated mode for colour 
images. Therefore, we evaluate different approaches for inpainting-based 
colour compression. Inpainting operators are able to reconstruct a large range
 of colours from a small colour palette of the known pixels. We exploit this 
 with a luma preference mode which uses higher sparsity in YCbCr colour 
 channels than in the brightness channel. Furthermore, we propose the first 
 full vector quantisation mode for an inpainting-based codec that stores only 
 a small codebook of colours. Our experiments reveal that both colour 
 extensions yield significant improvements.  
\end{abstract}


\section{Introduction}

In order to compress colour images, contemporary transform-based codecs have 
dedicated colour modes. For instance, JPEG \cite{PM92} and JPEG2000 
\cite{TM02} use chroma subsampling in YCbCr space, guided by the idea that 
structural information is visually more important than colour. Discarding some 
of the information of the chroma channels Cb and Cr allows them to store more 
accurate transform coefficients for a brightness representation of the image.
 
Inpainting-based approaches \cite{GWWB08,MMCB18,Pe19,SPMEWB14} rely on 
an entirely different concept. They create sparsity directly in the spatial 
domain. On colour images, these codecs usually rely on coarse quantisation of RGB 
values. Interestingly, inpainting operators tend to fill in gaps not only in 
the spatial domain, but also in the co-domain of colour values. Despite this 
property, little research has been invested into colour modes for 
inpainting-based codecs. In particular, the potential of techniques that 
create sparse colour palettes such as vector quantisation has remained 
unexplored so far.

\subsection{Our Contribution}
In the present paper, we introduce and evaluate different concepts for 
inpainting-based colour image compression. To this end, we propose two new 
colour modes for the recent RJIP codec which performs regular grid coding 
with joint inpainting and 
prediction \cite{Pe19}. Our first mode adapts a state-of-the-art 
colourisation-based 
concept \cite{PKW17} in YCbCr mode to the RJIP setting: We dedicate a higher 
budget to the image structure than to colour and augment it with efficient 
post-processing specifically tailored to fast Shepard interpolation 
\cite{Sh68}. Additionally, we propose the first full vector quantisation mode 
for inpainting-based compression. Our evaluation on the Kodak database 
\cite{Ko99} reveals that both colour modes outperform the original RGB mode 
significantly.

\subsection{Related Work}

While there are various inpainting-based codecs that can compress colour 
images \cite{DDI06,GWWB08,MMCB18,Pe19,SPMEWB14}, the only dedicated colour 
mode so far is the luma preference (LP) mode for the
R-EED codec \cite{PKW17}. It 
relies on the core idea to dedicate a higher budget to the luma channel of 
YCbCr space than to the colour channels. We adapt this concept to the setting 
of RJIP. In contrast to our approach, R-EED relies on a more complex 
inpainting method \cite{WW06} and a tree-based subdivision scheme to select 
and store positions of known data \cite{SPMEWB14}. In a broader sense, the LP 
mode resembles the chroma subsampling of JPEG \cite{PM92}.

Our second colour mode uses vector quantisation, a concept that  was already 
described by Shannon \cite{SW49} in his influential early work on information
 theory. Due to the large amount of research activity in the early 80s and 90s
 for compression of both visual and audio data, a full review is beyond the 
 scope of this work.  We refer the reader to the comprehensive monograph of 
 Gersho and Grey \cite{GG92} instead.

More recent works that deal with lossy compression and vector quantisation are
 rare. Venkateswaren and Ramana Rao \cite{VR07} quantise wavelet coefficients 
 from different sub-bands with vector clustering, while Somasundaram and Rani 
 \cite{SM12} focus solely on more efficient vector quantisation with a
  modified k-means clustering. For compression with neural networks, vector 
  quantisation is enjoying increasing popularity \cite{AMTC+17}.  However, 
  there it is not applied to colour values, but more generically to image 
  features or network parameters to be stored. Zhou et al. \cite{ZYCH19} 
  combine vector quantisation with inpainting, but in contrast to our work, 
  they quantise blocks of grey value data instead of individual colour values.

Even though they do not deal with colour, the work of Hoeltgen et 
al.~\cite{HPB18} comes close in spirit to our own research. They assess how 
clustering techniques affect inpainting-based reconstruction from sparse data.
However, they only use scalar quantisation. Their work confirms, together with
the findings of Celebi \cite{Ce11}, that the k-means clustering algorithm by
Lloyd \cite{Ll82} is one of the best clustering techniques for quantisation
. Consequentially, we rely on k-means clustering for our vector mode. We 
will discuss these methods in more detail in later sections.

\subsection{Organisation of the Paper} 
In Section \ref{sec:scalar}, we give a short review of the RJIP codec. 
Based on this foundation, we propose two novel colour extensions: 
a luma preference mode in Section 
\ref{sec:lpmode} and a vector mode in Section \ref{sec:vector}. Finally, 
we evaluate these new approaches in Section \ref{sec:experiments} and 
conclude our paper with a summary and an outlook on future work in Section 
\ref{sec:conclusion}.

\section{Review: Inpainting-based Compression\\ with Scalar Quantisation}
\label{sec:scalar}

\subsection{Inpainting}

At their core, all inpainting-based image compression codecs rely on 
interpolation from sparse image data. For a typical inpainting problem in RGB 
space, the colour image $\bm f: \Omega \rightarrow \R^3$ is only known at a 
few locations, the inpainting mask $K \subset \Omega$. The missing parts of 
the image domain $\Omega$ need to be reconstructed during decoding.

For the Shepard interpolation in RJIP \cite{Pe19}, computing an unknown pixel 
value $u_c(\boldsymbol{x}_i)$, $\bm x_i \in \Omega \setminus K$  for each 
channel $c \in \{R,G,B\}$ comes down to a simple weighted averaging of the 
known data according to 

\begin{equation}
\label{eq:shepard}
u_c(\boldsymbol{x}_i) = \frac{\sum_{\boldsymbol{x}_j \in K} 
w(\boldsymbol{x}_j - 
\boldsymbol{x}_i) f_c(\bm x_j)}{\sum_{\boldsymbol{x}_j \in K} 
w(\boldsymbol{x}_j - 
\boldsymbol{x}_i)}.
\end{equation}

Following Achanta et al. \cite{AAS17}, we use a truncated Gaussian $w$ 
with standard deviation $\sigma = \sqrt{(m \cdot n)/(\pi |K|)} $ for a 
discrete $m \times n$ image where $|K|$ is the number of mask pixels.

\subsection{Data Selection and Storage}

Each inpainting-based codec requires an adequate strategy 
to select and encode the inpainting mask.
RJIP stores the known data on a regular grid with grid size parameter $h$. 
This approach is fast and generates little overhead, but results in many 
colour values that need to be stored. RJIP compensates for this fact by 
\emph{joint inpainting and prediction}: While decompressing the image, it 
performs a partial inpainting whenever a new mask point is decompressed and 
uses this to predict the remaining mask points. Thus, only prediction errors 
need to be stored.

For the colour data corresponding to the mask positions, existing
inpainting-based codecs use \emph{uniform scalar quantisation} in each channel
: They map the 8 bit colour values to a reduced range range $\{0,\ldots,q-1\}$ 
by partitioning the tonal (i.e. colour value) domain into $q$ subintervals of 
equal length, limiting the amount of different colours for the known data to 
$q^3$. These quantised values are then stored with a suitable entropy encoder. 
RJIP relies on finite state encoding (FSE) \cite{Co14}, a fast coder similar 
to arithmetic coding.

For a given compression ratio, RJIP chooses the parameters $h$ and $q$ such 
that the best reconstruction quality for the desired file size is obtained.

\subsection{Tonal Optimisation}

RJIP benefits from \emph{tonal optimisation} in post-processing. For an image 
with $|K|$ mask points, it performs 
iterative random walks over all $3|K|$ R, G, or B values, adjusting them to a 
higher or lower quantisation level in case this yields a lower inpainting 
error. Even though this introduces a bias to the sparse stored data, it can 
increase the overall reconstruction quality in the large unknown areas 
significantly. We need to adapt this step for vector quantisation.

\section{RJIP with Luma Preference Mode}
\label{sec:lpmode}

\subsection{File Size Budget Distribution}

As a new colour mode for RJIP we consider a luma preference mode in
the YCbCr colour space. This technique
is an application of image colourisation that has been successfully used
in the R-EED codec \cite{PKW17}. Its core idea is to store more data for
the luma channel Y to reconstruct the image colour accurately and fill
in the colour information in the Cb and Cr channels from very sparse masks.

To this end, we distribute the total file size budget $B$ between
Y and CbCr according to a new parameter, the \emph{luma factor} $f$ 
such that $B_\text{Y} = f \cdot B_{\text{CbCr}}$. The compression pipeline
itself remains the same as in Section~\ref{sec:scalar}: On a regular mask, we
employ uniform quantisation, joint inpainting
and prediction, and FSE encoding. However, we use a separate
mask for the Y channel and a joint mask for the Cb and Cr channels.
The respective grid sizes $h$ and quantisation parameters $q$ are
chosen such that the budget constraints are fulfilled. 
Afterwards, we perform tonal optimisation for all channels.

\subsection{Tonal Optimisation}

 For tonal optimisation, we do not use the random walk trial-and-error 
 approach of RJIP described in Section \ref{sec:scalar}. Instead, we propose a 
 direct, more efficient algorithm. We exploit that the truncated weights in 
 Eq.~\eqref{eq:shepard} limit the influence of each pixel to a local area. In 
 the following, given a current pixel value 
 $u_i^\text{old}$ at $\bm x_i \in K$, 
 we want to find $u_i^\text{new}$, such that it minimises the mean squared 
 error (MSE).

To shorten notation, we define the value accumulation map $v$ as the numerator
 of Eq.~\eqref{eq:shepard} and write $w_{i,j}$ for $w(\bm x_j - \bm x_i)$ for 
 weights at positions $\bm x_j$ from a neighbourhood  $\mathcal{N}_i$ relative
  to its centre $\bm x_i$. Then the new error after changing $u_i^\text{old}$ 
  to $u_i^\text{new}$ is given by
\begin{equation}
e (u_i^\text{new})  = \sum_{\bm x_j \in \mathcal{N}_i }
\left(f_j - \frac{v_j + w_{i,j} \left(u_i^\text{new} - 
	u_i^\text{old}\right)}
{w_j}
\right)^2,
\end{equation}
where $$w_j = \sum_{\bm x_j \in \mathcal{N}_i } w_{i,j}.$$
Since the optimal new tonal value $u_i^\text{new}$ should minimise $e(\cdot)$, 
we solve $\frac{d}{d u_i^\text{new}} e(u_i^\text{new})=0$ 
for $u_i^\text{new}$ and obtain
\begin{equation}
u_i^\text{new} = \frac{\sum_{\bm x_j \in \mathcal{N}_i } 
	\frac{G_{i,j}}{w_j}
	\left(f_j - \frac{v_j - G_{i,j} u_i^\text{old}}{w_j}\right)}
{\sum_{\bm x_j \in \mathcal{N}_i}
	\frac{G_{i,j}^2}{w_j^2}} \, .
\end{equation}

Instead of testing neighbouring quantisation levels, we can now compute the 
locally optimal value directly. However, note that we need to project these 
unconstrained solutions onto the set of admissible quantisation values after 
each computation. Iterating these steps eventually converges to a fully 
optimised inpainting mask. 

\section{RJIP with Vector Quantisation}
\label{sec:vector}

As a second alternative to the LP mode from Section \ref{sec:scalar}, we 
propose the
first full vector quantisation mode for inpainting-based colour image 
compression. In the following we describe the corresponding modifications to 
the compression pipeline of RJIP.

\subsection{Clustering with k-means}
\label{sec:kmeans}

\begin{figure}[t]
	\centering
	\tabcolsep1pt
	\begin{tabular}{cc}
	image & histogram \\ 
	\includegraphics[height=2.7cm]{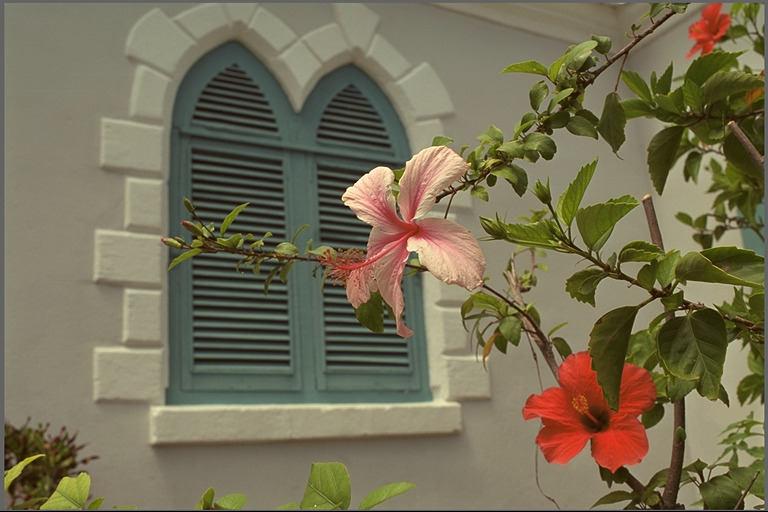} & 
	\includegraphics[height=2.7cm]{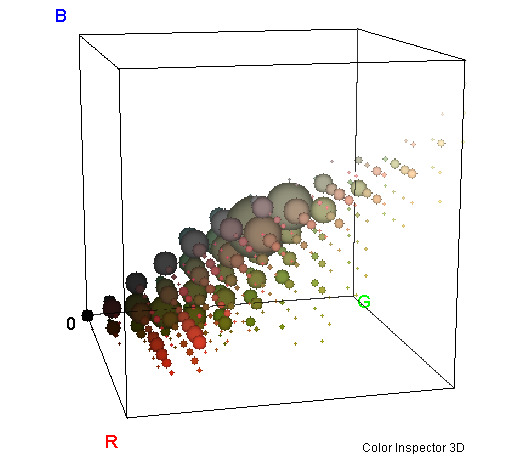}\\[-0.5mm]
	(a) original \emph{kodim07} & \\[1mm]
	\includegraphics[height=2.7cm]{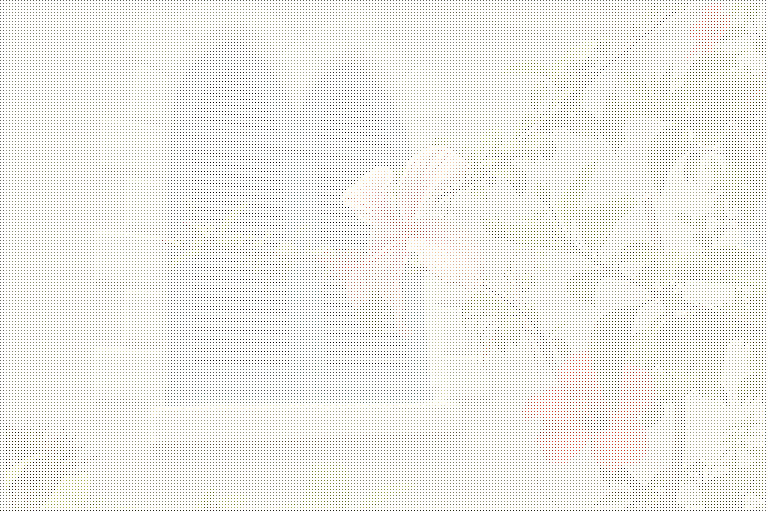}&
	\includegraphics[height=2.7cm]{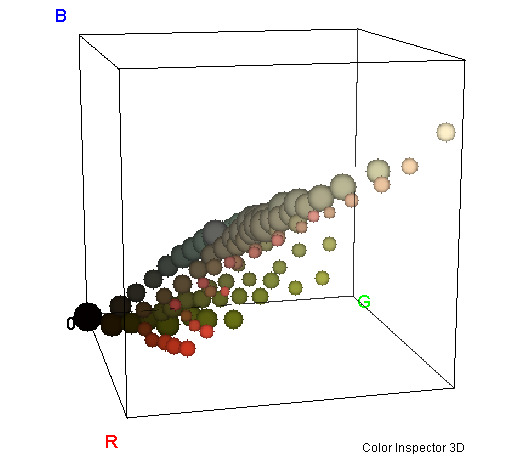}\\[-0.5mm]
	(b) inpainting mask &  \\[1mm]
	\includegraphics[height=2.7cm]{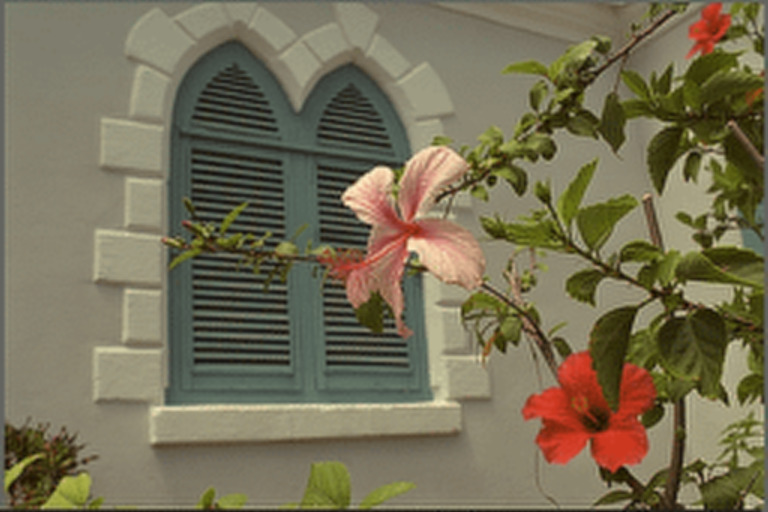} &
	\includegraphics[height=2.7cm]{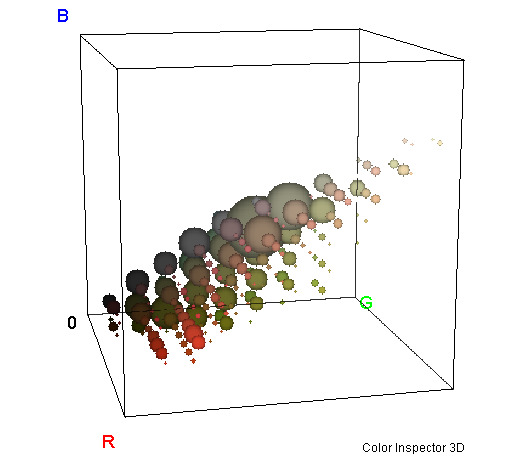} \\[-0.5mm]
	(c) reconstruction &  \\
	\end{tabular}
	\caption{Colour histograms for the original image \emph{kodim07} 
		\cite{Ko99} and the 
		reconstructed image with Shepard inpainting.  The histograms (created 
		with Colour Inspector 3D \cite{Ba04})  show each bin as a ball with 
		radius proportional to the number of contained colours. Inpainting can
		 reconstruct a good approximation to the original histogram from only 
		 few quantised colours present in the inpainting mask. Here, the image
		  is compressed to a ratio of 50:1 and the mask has only 92 different 
		  colours  \label{fig:kodakhist}
	}
\end{figure}


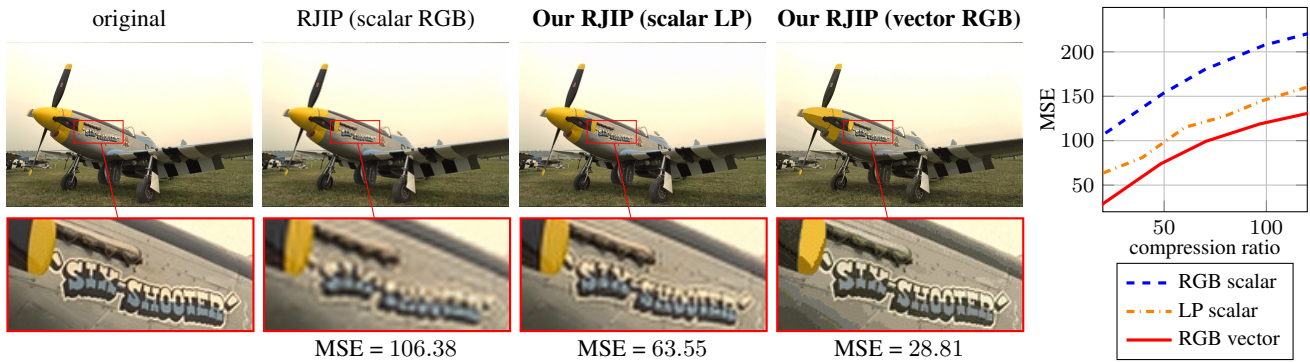
\begin{figure*}[t]
	\centering
	\tabcolsep-1pt
	\noindent\begin{minipage}[t]{0.75\linewidth}
		 \vspace{0pt}
	\begin{tabular*}{1\linewidth}{@{\extracolsep{\fill}}cccc}
		original & RJIP (scalar RGB) & \textbf{Our RJIP (scalar LP)}  & 
		\textbf{Our RJIP (vector RGB)} \\
		\begin{tikzpicture}[spy using outlines={rectangle,red,ultra 
			thick,magnification=5,width=0.245\linewidth,  height=1.5cm, 
			connect spies}]
		\node 
		{\pgfimage[width=0.245\linewidth]{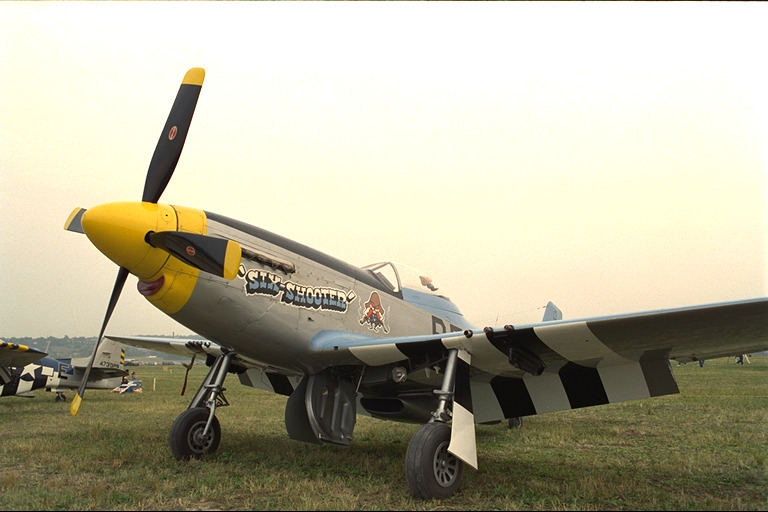}};
		\spy on (-0.41,-0.1) in node [left] at (1.65,-2);
		\end{tikzpicture}
		&
		\begin{tikzpicture}[spy using outlines={rectangle,red,ultra 
			thick,magnification=5,width=0.245\linewidth,  height=1.5cm, 
			connect 
			spies}]
		\node 
		{\pgfimage[width=0.245\linewidth]{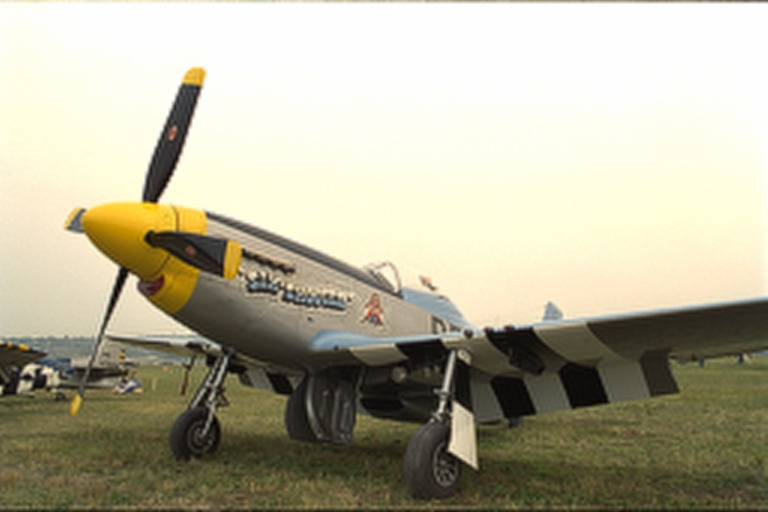}};
		\spy on (-0.41,-0.1) in node [left] at (1.65,-2);
		\end{tikzpicture}
		&
		\begin{tikzpicture}[spy using 
		outlines={rectangle,red,ultra 
			thick,magnification=5,width=0.245\linewidth,  height=1.5cm, 
			connect 
			spies}]
		\node 
		{\pgfimage[width=0.245\linewidth]{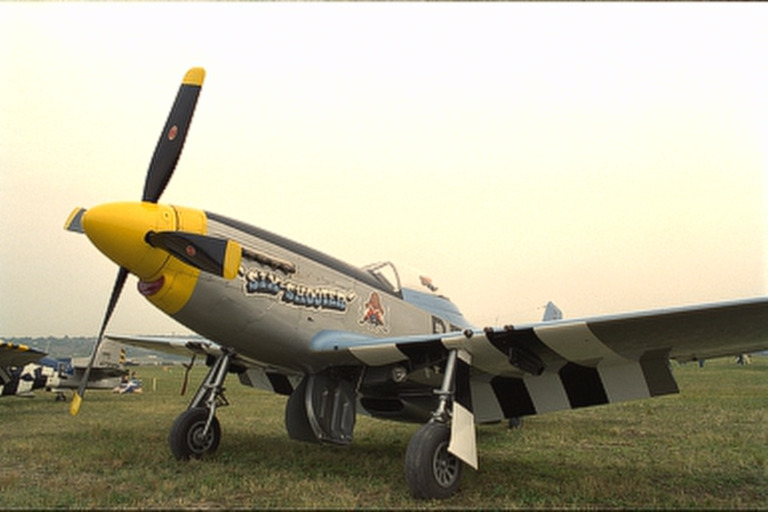}};
		\spy on (-0.41,-0.1) in node [left] at (1.65,-2);
		\end{tikzpicture}
		&
		\begin{tikzpicture}[spy using outlines={rectangle,red,ultra 
			thick,magnification=5,width=0.245\linewidth,  height=1.5cm, 
			connect 
			spies}]
		\node 
		{\pgfimage[width=0.245\linewidth]{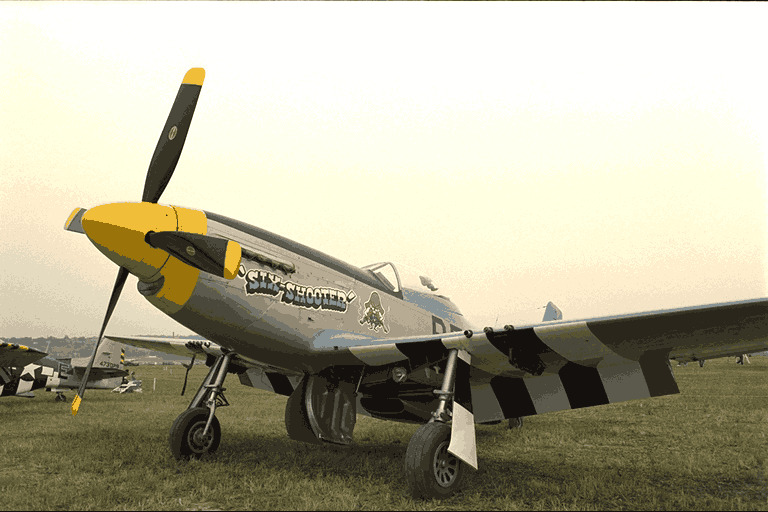}};
		\spy on (-0.41, -0.1) in node [left] at (1.65,-2);
		\end{tikzpicture} 
		\\[-0.5mm]
		& MSE~=~$106.38$
		& MSE~=~$63.55$
		&MSE~=~$28.81$
		\\
	
	\end{tabular*}
		\end{minipage}%
\begin{minipage}[t]{0.25\linewidth}
	 \vspace{0pt}
	 \centering		
	\begin{tikzpicture}
	\begin{axis}[height=4.3cm, width=4.3cm,
	ymin=20, ymax=250, xmin=20, xmax=120,
	xtick = {50,100},
	ytick = {50,100,150,200},
	label style={font=\footnotesize},
	tick label style={font=\footnotesize},  
	xlabel={compression ratio},
	ylabel={MSE},
	ylabel style={yshift=-0.14cm},
	xlabel style={yshift=0.14cm},
	grid = major,
	legend entries={RGB scalar, LP scalar, 
		RGB vector},
	legend cell align=left,
	legend 
	style={at={(0.5,-0.25)},anchor=north,font=\footnotesize}]
	\addplot+[mark=none, dashed, very thick, blue] table 
	{data/kodim20_scalar_residuals.tab};
	\addplot+[mark=none,dashdotted, very thick, orange] table[x index=0, y 
	index=1] 
	{data/kodim20_scalar_luma.csv};
	\addplot+[mark=none, very thick,red] table[x 
	index=0, y index=1] {data/kodim20_vector_pixels.tab};
	\end{axis}
	\end{tikzpicture}
		\end{minipage}
		
	\caption{\label{fig:rjipvector} RJIP compression results (RGB-MSE)
		for \emph{kodim20} ($768 \times 512$ pixels) with a compression ratio 
		of 20:1. Both our colour modes outperform RGB RJIP significantly.
		On images with low and moderate amount of texture, vector quantisation
		also outperforms LP mode considerably.
	}
\end{figure*}

\begin{figure*}[t]
	\centering
	\tabcolsep-1pt
	\noindent\begin{minipage}[t]{0.75\linewidth}
		 \vspace{0pt}
	\begin{tabular*}{1\linewidth}{@{\extracolsep{\fill}}cccc}
		original & RJIP (scalar RGB) & \textbf{Our RJIP (scalar LP)}  & 
		\textbf{Our RJIP (vector RGB)} \\
		\begin{tikzpicture}[spy using outlines={rectangle,red,ultra 
			thick,magnification=3,width=0.245\linewidth,  height=1.5cm, 
			connect spies}]
		\node 
		{\pgfimage[width=0.245\linewidth]{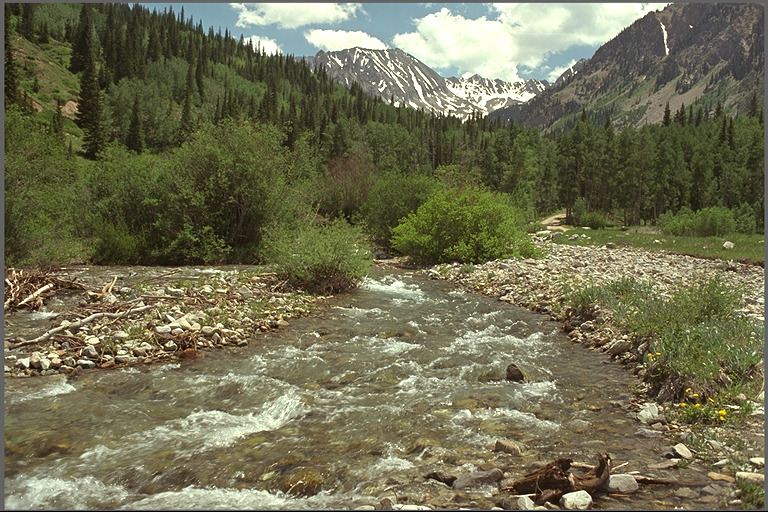}};
		\spy on (-0.91,-0.05) in node [left] at (1.65,-2);
		\end{tikzpicture}
		
		&
		\begin{tikzpicture}[spy using outlines={rectangle,red,ultra 
			thick,magnification=3,width=0.245\linewidth,  height=1.5cm, 
			connect 
			spies}]
		\node 
		{\pgfimage[width=0.245\linewidth]{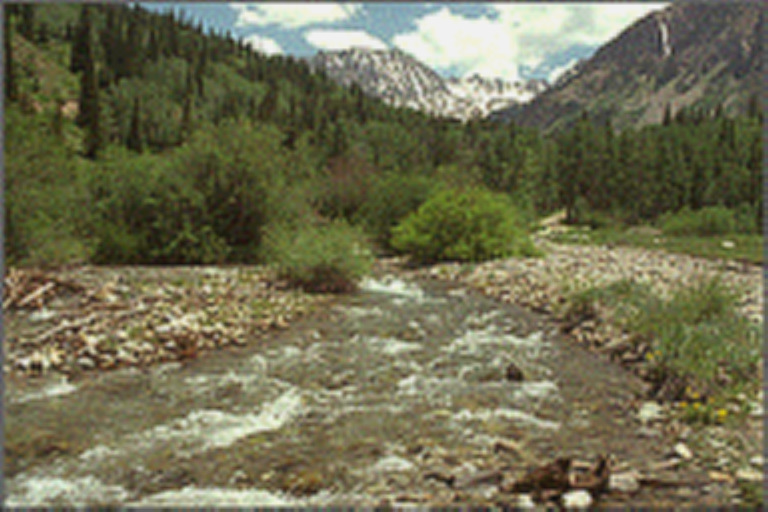}};
		\spy on (-0.91,-0.05) in node [left] at (1.65,-2);
		\end{tikzpicture}
		&
		\begin{tikzpicture}[spy using 
		outlines={rectangle,red,ultra 
			thick,magnification=3,width=0.245\linewidth,  height=1.5cm, 
			connect 
			spies}]
		\node 
	{\pgfimage[width=0.245\linewidth]{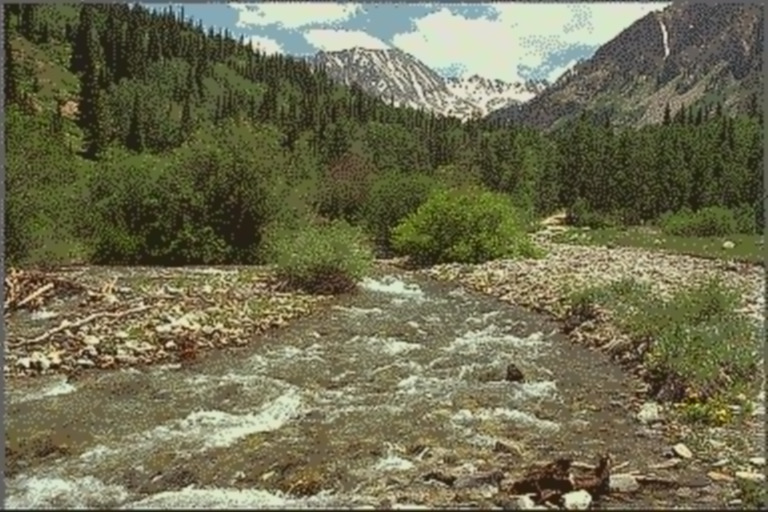}};
		\spy on (-0.91,-0.05) in node [left] at (1.65,-2);
		\end{tikzpicture}
		&
		\begin{tikzpicture}[spy using outlines={rectangle,red,ultra 
			thick,magnification=3,width=0.245\linewidth,  height=1.5cm, 
			connect 
			spies}]
		\node 
		{\pgfimage[width=0.245\linewidth]{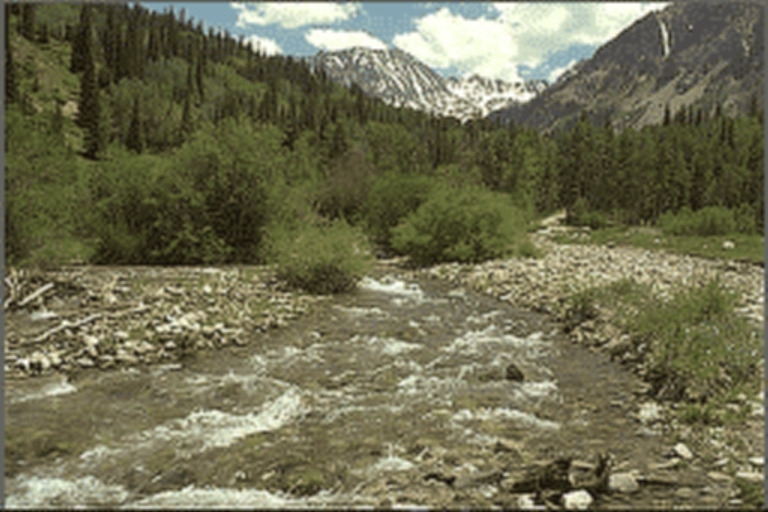}};
		\spy on (-0.91,-0.05) in node [left] at (1.65,-2);
		\end{tikzpicture} 
		\\[-0.5mm]
		& MSE~=~$580.98$
		& MSE~=~$398.64$
		&MSE~=~$462.47$
		\\
		
	\end{tabular*}
		\end{minipage}%
\begin{minipage}[t]{0.25\linewidth}
	 \vspace{0pt}
	 \centering		
	\begin{tikzpicture}
	\begin{axis}[height=4.3cm, width=4.3cm,
	ymin=250, ymax=700, xmin=20, xmax=120,
	xtick = {50,100},
	ytick = {300,400,500,600,700},
	label style={font=\footnotesize},
	tick label style={font=\footnotesize},  
	xlabel={compression ratio},
	ylabel={MSE},
	ylabel style={yshift=-0.14cm},
	xlabel style={yshift=0.14cm},
	grid = major,
	legend entries={RGB scalar, LP scalar, 
		RGB vector},
	legend cell align=left,
	legend 
	style={at={(0.5,-0.25)},anchor=north,font=\footnotesize}]
	\addplot+[mark=none, dashed, very thick, blue] table 
	{data/kodim13_scalar_residuals.tab};
	\addplot+[mark=none,dashdotted, very thick, orange] table[x index=0, y 
	index=1] 
	{data/kodim13_scalar_luma.csv};
	\addplot+[mark=none, very thick,red] table[x 
	index=0, y index=1] {data/kodim13_vector_pixels.tab};
	\end{axis}
	\end{tikzpicture}
		\end{minipage}
		
	\caption{\label{fig:rjipscalar} RJIP compression results (RGB-MSE)
		for \emph{kodim13} ($768 \times 512$ pixels) with a compression ratio 
		of 50:1. Both our colour modes outperform RGB RJIP significantly.
		On images with a high amount of texture, LP mode
		outperforms vector quantisation considerably.
	}
\end{figure*}
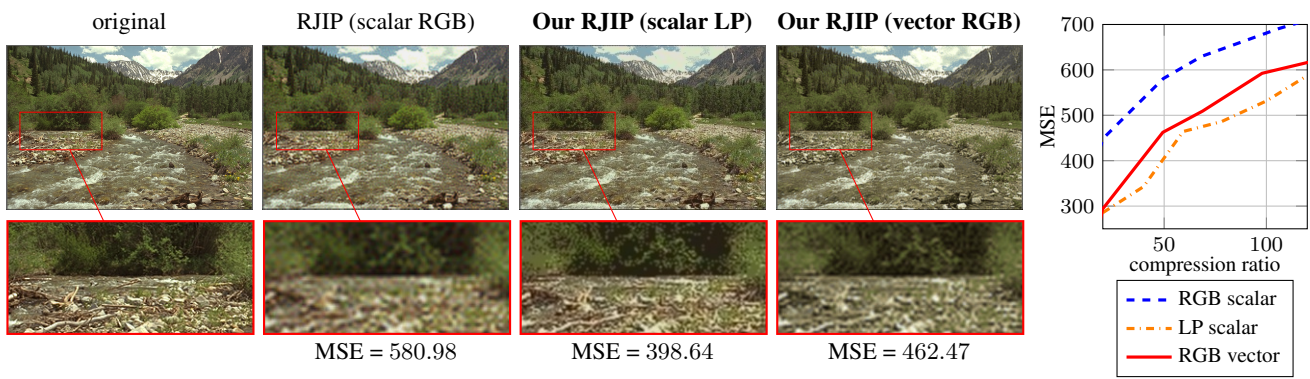

Experimental evaluations have shown that the simple 
k-means algorithm for clustering by Lloyd \cite{Ll82} is still
one of the most successful techniques for vector quantisation \cite{Ce11}. 
Given an initial set of colour vectors $V:=\{\bm v_1,...,\bm v_n\}$, k-means 
clustering aims to partition
$V$ into $k$ disjoint clusters $C_1,...,C_k \subset V$ such that they minimise 
the cumulative squared Euclidean distance of the points in each cluster 
$C_\ell$ to the corresponding cluster centre $\bm \mu_\ell$.

To achieve this goal, the k-means algorithm first randomly selects $k$ cluster
 centres.

Alternating assignment and update steps refine this initialisation iteratively
. The assignment step maps all colours to the cluster with the nearest mean 
value. In the subsequent update, the mean values are set to the centroids of 
the newly assigned clusters. Note that there are many more sophisticated 
initialisation strategies than the random one, e.g.~k-means++ \cite{AV07}, 
histogram-based approaches \cite{GC14,Te90,Si86,FD81,Sc79}, and iterative 
subdivision methods \cite{SD04}. Even though the initialisation method impacts 
the quantisation error, we experienced empirically that it does not impact the 
final compression performance significantly. The tonal and cluster 
post-processing steps described in the following paragraphs ensure that we 
achieve good final results regardless of the initialisation.

A representative histogram of an image after vector quantisation can be seen 
in Fig.~\ref{fig:kodakhist} (c). The same figure also reveals that inpainting 
can approximate such a histogram well from data that is simultaneously sparse 
in the spatial domain and the colour space after coarse vector quantisation.

\subsection{Cluster Post-Processing}

For tonal optimisation, we use our improved method from Section 
\ref{sec:lpmode} to find optimal unconstrained values. After each iteration, 
we project them onto the closest cluster centre w.r.t.~Euclidean distance. 
However, for vector quantisation we consider an additional post-processing 
step.

Just as the original colour values do not necessarily yield the best inpainting 
result, the k-means clusters are not necessarily optimal w.r.t.~reconstruction 
quality. As another post-processing step, we can check if moving a cluster 
centre to a neighbouring vector from $\{0,...,256\}^3$ decreases the MSE and 
each such global change affects all known pixels with the same label. Thus, 
we optimise the quantisation codebook for final reconstruction quality and can
 simultaneously compensate for suboptimal initialisations of the k-means 
 algorithm.  Even though the overall quantitative gain is negligible for lower
  compression ratios, it becomes significant for higher compression ratios.

\subsection{Storing Colour Palettes and Entropy Coding}

In contrast to uniform scalar quantisation, we need to store the colour 
palette used by the encoder. In the header we first save the number 
$q \leq 256$ of quantised colours with 1 byte and the cluster centres 
themselves with 1 byte per channel. The position of known data points is 
stored as in standard RJIP, but their value is now represented by a cluster 
centre index ($0,...,q-1$) from the stored codebook.

Finally, we apply entropy coding. Unfortunately, the joint prediction and 
inpainting from Section \ref{sec:scalar} yields unsatisfactory results in the 
vector case. While inpainting can still predict the colour value of 
neighbouring known data, the codebook labels are not tied to the distance 
between vectors in the colour space. Thus, an accurate prediction of the 
colour itself can still yield a high error in the label prediction. Therefore,
 we replace this step with Prediction by Partial Matching (PPM) \cite{CW84}.

\begin{figure}[t]
	\centering
	\tabcolsep0pt
	\begin{tikzpicture}
	\begin{axis}[height=5cm, width=8cm,
	ymin=20, ymax=300, xmin=20, xmax=120,
	xtick = {10,30,50,70,90,110,130,150},
    ytick = {10,50,90,130,170,210,250},
	xlabel={compression ratio},
	ylabel={RGB MSE},
	grid = major,
	legend entries={RJIP-RGB (scalar), RJIP-LP (scalar), 
	RJIP-RGB (vector)},
	legend cell align=left,
 legend style={legend pos=south east, font=\footnotesize},]

	\addplot+[mark=none, dashed, very thick, blue] table[x index=0, y index=1, 
	col sep = comma] {data/kodak_scalar_residuals.csv};
	\addplot+[mark=none, dashdotted, very thick, orange] table[x index=0, y 
	index=1] {data/kodak_rjip_luma_2.csv};
	\addplot+[mark=none, very thick, red] table[x index=0, y index=1, col 
	sep = comma] {data/kodak_vector_pixels.csv};

	\end{axis}
	
	\end{tikzpicture}
	\caption{\label{fig:rjip} On the full Kodak database, both our new colour 
		modes for RJIP outperform RGB scalar RJIP consistently and yield a 
		similar error on average.}
\end{figure}
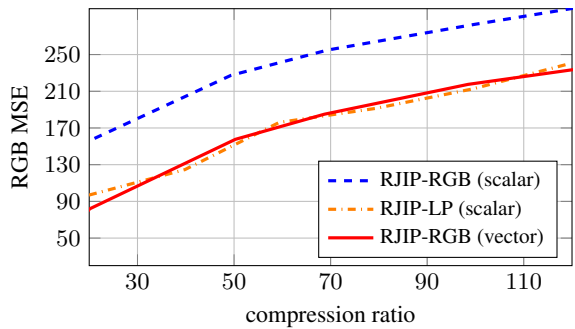

The original version of PPM uses the linear sequence of previously encoded 
symbols contexts for deriving conditional probabilities for entropy coding. 
However, this approach discards the 2-D relations of image pixels. Instead, we
 iterate over all mask pixel and use their three direct neighbours that have 
 already been encoded as a 2-D context. The resulting conditional 
 probabilities are then encoded with arithmetic coding \cite{Ri76}.

\section{Experiments}
\label{sec:experiments}

On the well-known Kodak image database \cite{Ko99}, we compare the RGB scalar 
mode of the original RJIP \cite{Pe19} to our two new colour modes: RGB with 
vector quantisation and YCbCr luma preference mode (LP) with scalar
 quantisation. We use the MSE over all RGB channels as our error measure. For 
 LP mode, we choose the luma factor $f \in \{0.5,0.6,0.7,0.8,0.9\}$ that 
 minimises the MSE. 

Fig.~\ref{fig:rjip} shows that both our new colour modes for RJIP consistently
 outperform the scalar RGB mode by a large margin. This also results in a 
 significant improvement of visual quality, which is illustrated by 
 Fig.~\ref{fig:rjipvector} and Fig.~\ref{fig:rjipscalar}. On average over the 
 whole database, both colour modes yield a similar error. However, a more 
 detailed analysis shows that both modes have distinct advantages on different
  types of image content.

For images with low amounts of texture, vector quantisation is the better 
choice and can reduce the MSE by up to 54\%  compared to scalar LP as shown in
 Fig.~\ref{fig:rjipvector}. This yields a noticeable increase in visual 
 quality due to higher RGB mask densities. However, on heavily textured images
  such as the one in Fig.~\ref{fig:rjipscalar}, LP mode yields better results.
   Here, more data in the luma channel allows a more accurate reconstruction 
   of the image structure. 

In general, vector quantisation is computationally faster than the LP 
mode. Even though vector quantisation is more complex than uniform scalar 
quantisation due to k-means clustering, it does not require to optimise the 
luma factor $f$. This yields a runtime reduction of up to 50\%.

\FloatBarrier
\section{Conclusion}
\label{sec:conclusion}

Our two new colour modes for RJIP offer a significant visual and quantitative
improvement over the standard RGB mode. Our evaluation of the vector 
quantisation mode reveals
that inpainting-based image compression can reconstruct a wide range of 
colours from a sparse codebook. On highly textured images, a scalar luma 
preference mode can reproduce the image structure more accurately. In future 
research we want to investigate our colour modes on data with many channels, 
e.g. hyperspectral imaging \cite{ASPW17}.

\FloatBarrier

\bibliographystyle{IEEEbib}
\bibliography{bib}

\end{document}